\documentstyle[12pt,leqno]{article}%% Stile articolo numero formule a sinistra
 \newcommand{\beq}{\begin{equation}}                       % Abbreviare
 \newcommand{\eeq}{\end{equation}}                         % la scrittura
 \newcounter{nt}[section]                                  % Num sez
 \newcounter{nl}[section]                                  % Num sez
 \date{}                                                   % No data

\begin{document}

 \title{ \bf Parabolic - hyperbolic boundary layer
 }

\vspace{15mm}

\author{\sc Monica De Angelis
   \thanks{ Facolt\`{a} di
             Ingegneria, Dipartimento di Matematica e Applicazioni, via Claudio 21, 80125, Napoli.}}

 \maketitle

%\end{document}
 \begin{abstract}
A  boundary value  problem related to a
parabolic higher order operator
with a small parameter $\varepsilon  $ is analized.  For $\varepsilon$
tends to zero, the reduced operator is hyperbolic. When $t$ $\rightarrow
\infty$ and $\varepsilon $   $\rightarrow
0$ a parabolic hyperbolic boundary layer appears. In this paper a
rigorous asymptotic approximation uniformly valid for all $t$ is
established.
 \end{abstract}

  \vspace{7mm}

 \section{Introduction}
 \setcounter{equation}{0}

 \hspace{5.1mm}

The parabolic operator
\vspace{3mm}

\beq                     \label{11
}
 {\cal L}_ \varepsilon = \partial_{xx} ( \varepsilon \partial _{t}+
 c^2) - \partial _{tt}
 \eeq
\vspace{3mm}
\noindent
is related to the well known Kelvin -Voigt viscoelastic model.
 Further, it characterizes also the principal part of numerous models
 with non linear dissipation, such as

\beq                     \label{12}
 {\cal L}_ \varepsilon =\beta(u,u_x,u_t).
 \eeq
\vspace {3mm}
Tipical example is the perturbed Sine Gordon equation.\cite{ddr} Moreover, by
means

of (\ref{12}), wave equations with non linear terms are
regularized obtaining a priori estimates and considering the
$\varepsilon$ parameter vanishing.\cite{kl}. Further  third operators
are also considered to value the Cauchy problem for a second order
hyperbolic equation \cite{mm} or to regularizate parabolic forward-
backward equations.\cite{bb}

Singular perturbation problem related to equations like (\ref{12}) have
interest also to evaluate the influence of the dissipative causes on
the wave propagation. \cite{n}. In particular, in the linear case
$\beta=f(x,t)$, it`s interesting to compare the effects of the diffusion
with the pure waves which occur when $\varepsilon =0$ . In this case one
has a parabolic- hyperbolic boundary layer with the unique singularity
for $t \rightarrow \infty.$

In this paper , we consider the strip problem for equation (\ref{12})
and analyze the singular perturbation problem when $\beta =f(x,t)$ is
linear. The Green function related to this problem has been already
determined in term of a rapidly decreasing Fourier series.\cite {mda}.

An appropriate analysis of this series when $\varepsilon \rightarrow 0$
allows to obtain a rigorous asymptotic estimate of the solution,
uniformly valid even  $t \rightarrow \infty.$

\section{Statement of the problem}
 \setcounter{equation}{0}

 \hspace{5.1mm}

If $v(x,t)$ is a function defined in
\vspace{4mm}

$\ \ \ \ \ \ \ \ \ \ \ \  \ \ \ \ \ \ \   \Omega =\{(x,t) : 0 < x <
l, \  \ t \geq 0 \},$

\vspace{3mm}
\noindent
with  $l $ arbitrary positive constant, let IBC the following system of
initial-boundary conditions:

\vspace{3mm}
  \beq                                                     \label{21}
  \left \{
   \begin{array}{ll}
   & v(x,0)=f_0(x), \  \    v_t(x,0)=f_1(x), \  \ x\in [0,l],\vspace{2mm}  \\
    & v(0,t)=\psi_0, \  \ v(l,t)=\psi_1, \ \ \  \ t\geq 0,
   \end{array}
  \right.
 \eeq

\vspace{3mm}\noindent
with $f_i, \psi_i$ ($i=0,1)$ regular data.

Consider the operators:

\beq                     \label{22}
 {\cal L}_ 0 = c^2 \partial{xx}- \partial{tt} ; \ \  {\cal
 L}_\varepsilon = {\cal L}_0 +\varepsilon \partial{xxt}
 \eeq
\vspace {3mm}
\noindent
and denote by $u_0$ and $u_\varepsilon$ the solutions of the problems:
\noindent

$ Problem P_0 : \ \ \ \ \  {cal L}_0 u_0= -f$  with IBC {\ref(21)}

\noindent

$Problem  P_\varepsilon : \ \ \ \ \ {cal L}_\varepsilon u_\varepsilon=
-f $ with IBC {\ref(21)},

\noindent
where $ f(x,t)$ is a prefixed source term.

To obtain a rigorous approximation of $ u_\varepsilon$ when$ \varepsilon
\rightarrow 0$, we put
\vspace{3mm}
\beq                                         \label{23}
u(x,t,\varepsilon)=  u_0(x,t) + \varepsilon r(x,t,\varepsilon)
\eeq
\vspace{3mm}
\noindent
where $u_0$ is the well-known solution of the classical problem $P_o$,
while the error term represent the solution of the $Problem P_r$:

\vspace{3mm}
  \beq                                                     \label{24}
  \left \{
   \begin{array}{ll}
   & {cal L} \varepsilon r = - F(x,t)   \ \ \ \ (x,t) \in \Omega \\
   & v(x,0)=f_0(x), \  \    v_t(x,0)=f_1(x), \  \ x\in [0,l],\vspace{2mm}  \\
    & v(0,t)=\psi_0, \  \ v(l,t)=\psi_1, \ \ \  \ t\geq 0,
   \end{array}
  \right.
 \eeq

\vspace{3mm}\noindent
with $ F(x,t) = \partial{xxt} u_0$. Therefore, following results in
\ref{mda},one has:
\vspace {3mm}
\beq                                      \label{25}
r(x,t,\varepsilon)= -\int_{0}^{l} d\xi \\ \int_{0}^{t}
F(\xi,\tau,\varepsilon) \\\\\\ G(x,\xi,t-\tau) \ d\tau
\eeq

\vspace{3mm}\noindent
where $G(x,\xi,t)$ is the Green function related to $ {\cal
L}_\varepsilon$ operator.

In particular, for all integer $n \geq 1$ , letting:
\vspace{3mm}
\beq                 \label{26}
\gamma_n =\frac{\pi}{l} n\ \ \ a_n = \frac{\varepsilon}{2} \gamma_n
^2 \ \ \\ k=\frac{2cl}{\pi\varepsilon} \
\eeq
\hspace* {2cm} \[ b_n= \gamma_nc \sqrt{1-(n/k)^2} \ \  H_n = e^{/a_n}
\frac{sen (b_n t)}{b_n}, \]
\vspace{3mm}\noindent
i

\beq                                     \label{26}
G(x,\xi,t)=\frac{2}{l} \ \ \\ \ \sum_{n=1}^{\infty} \ \ H_n(t) \ \
\sin \gamma_n x  \  \sin \gamma_n \xi
\eeq

\vspace {3mm}\noindent
with
\vspace{3mm}
\beq                             \label{27}
H_n(t)= \ \ \frac{e^{-bn^{2}t}}{bn^2 \sqrt{1-(k/n)^2}} \
\sinh(bn^2t \sqrt{1-(k/n)^2})
\eeq

\vspace{3mm}
\noindent
and
\vspace{3mm}
\beq                 \label{28}
b= \frac{\pi^2}{2l^2}\varepsilon = q \varepsilon, \\\\ \ \ \ \ k=\frac{2cl}{\pi\varepsilon} \
\ \ \ \\\ \  \  \gamma_n =\frac{\pi}{l} n.
\eeq

Now, denote with $u(x,t)$ the solution of the reduced problem obtained
by (\ref{21}) with $\varepsilon=0$.
To obtain an asymptotic approximation for $w(x,t) $ when $\varepsilon
\rightarrow 0$, we put:

\vspace{3mm}
\beq                                         \label{22}
w(x,t,\varepsilon)= e^{-\varepsilon t}  u(x,t) + r(x,t,\varepsilon)
\eeq
\vspace{3mm}
\noindent
where  the error $r(x,t,\varepsilon )$ must be evaluated.

By means of standard computations one verifies that $r(x,t,\varepsilon)$
is the solution of the problem:

onsider the operators
\vspace{3mm}
  \beq                                                     \label{23}
  \left \{
   \begin{array}{ll}
    & \partial_{xx}(\varepsilon
r_{t}+c^2 r) - \partial_{tt}r=f(x,t,\varepsilon),\ \  \
       (x,t)\in D,\vspace{2mm}\\
   & r(x,0)=0, \  \    r_t(x,0)=0, \  \ x\in [0,l],\vspace{2mm}  \\
    & r(0,t)=0, \  \ r(l,t)=0, \  \ 0<t < T,
   \end{array}
  \right.
 \eeq

\vspace{3mm}\noindent
where the source term $f(x,t,\varepsilon)$ is:

\vspace {3mm}

\beq                                                \label{24}
f(x,t,\varepsilon)=F(x,t)(1-e^{-\varepsilon t})+ \ e^{-\varepsilon t}[-\varepsilon
\lambda_t+\varepsilon^2 (u+u_{xx})]
\eeq

\vspace{3mm} \noindent
with  $\lambda= 2u+u_{xx}$.

The problem (\ref{23}) has already been solved in \cite{mda} and
the solution is given by:

\vspace {3mm}
\beq                                      \label{25}
r(x,t,\varepsilon)= -\int_{0}^{l} d\xi \\ \int_{0}^{t}
f(\xi,\tau,\varepsilon) \\\\\\ G(x,\xi,t-\tau) \ d\tau
\eeq

\vspace{3mm}\noindent

where $G(x,\xi,t)$ is:

\vspace{3mm}

\beq                                     \label{26}
G(x,\xi,t)=\frac{2}{l} \ \ \\ \ \sum_{n=1}^{\infty} \ \ H_n(t) \ \
\sin \gamma_n x  \  \sin \gamma_n \xi
\eeq

\vspace {3mm}\noindent
with
\vspace{3mm}
\beq                             \label{27}
H_n(t)= \ \ \frac{e^{-bn^{2}t}}{bn^2 \sqrt{1-(k/n)^2}} \
\sinh(bn^2t \sqrt{1-(k/n)^2})
\eeq

\vspace{3mm}
\noindent
and
\vspace{3mm}
\beq                 \label{28}
b= \frac{\pi^2}{2l^2}\varepsilon = q \varepsilon, \\\\ \ \ \ \ k=\frac{2cl}{\pi\varepsilon} \
\ \ \ \\\ \  \  \gamma_n =\frac{\pi}{l} n.
\eeq
\vspace{1cm}
 \section{Analysis of $G(x,t, \xi,\varepsilon)$ when $\varepsilon$ tends to
 zero.}
 \setcounter{equation}{0}

 \hspace{5.1mm}
In order to investigate the behaviour of the Green function G when
parameter $\varepsilon \rightarrow 0 $, referring to the function G
defined in (\ref{26}), let:

\vspace{3mm}
\beq                             \label{31}
H^1_n(t)= \ \ \frac{e^{-bn^{2}t}}{bn^2 \sqrt{(k/n)^2-1}} \
\sin{bn^2t \sqrt{(k/n)^2-1}}
\eeq

\vspace {3mm}\noindent

and
\vspace{3mm}
\beq                             \label{32}
G(x,\xi,t)= \frac{2}{l} \ \ \\ \{ \sum_{n=1}^{[k]} \  H^1_n(t) \
 \\  +  \sum_{[k]+1}^{\infty} \\ H_n(t) \}
\sin \gamma_n x \ \sin \gamma_n \xi \ = \ G_1 + G_2.
\eeq

\vspace{3mm}
If $\alpha $ is an arbitrary constant such that:
\vspace{3mm}
\beq                         \label{33}
1/2 < \alpha <1 , \ \ \ \ \  \bar n
= \frac{2cl}{\pi \varepsilon ^\alpha},\ \ \ \
\eeq

\vspace {3mm} \noindent
the term $G_1 $ of $G $  can be given the forms:
\vspace {3mm}
\beq                       \label{34}
G_1(x,\xi,t)= \frac{2}{l} \ \ \\ \ \{ \sum_{n=1}^{[\bar n]} \  H^1_n(t) \
 \\  +  \sum_{[\bar n]+1}^{[k]} \\ H^1_n(t) \}
\sin \gamma_n x  \ \ \sin \gamma_n \xi \ .
\eeq
 \vspace{3mm}\noindent
It is easy to prove that if $1\leq n \leq [\bar n]$ it holds:
\vspace{3mm}

\beq                 \label{35}
\sqrt{(k/n)^2-1} \geq
\frac{\sqrt{1-\varepsilon^{2(1-\alpha)}}}{\varepsilon^{1-\alpha}}; \ \ \
 e^{-bn^{2}t} \leq  e^{-qt \varepsilon}.
\eeq

\vspace{3mm}\noindent
Otherwise, if  $[\bar n]+1 \leq n \leq [k] $:
\vspace{3mm}

\beq                 \label{36}
\sqrt{(k/n)^2-1} \geq
\frac{ \sqrt{\pi\varepsilon \beta } \sqrt{4cl-\beta \pi \varepsilon}}{
 (2cl-\pi \varepsilon\beta)}; \ \ \
 e^{-bn^{2}t} \leq  e^{-2 c^2 t / \varepsilon^{2\alpha-1}},
\eeq

\vspace{3mm}\noindent
where $0<\beta<1$. In particular, if $k$ is an integer we will assume
$\beta=1$ and we will explictly consider the term with $n=k$, having $t
e^{-2c^2 t/\varepsilon}$.

\vspace{3mm}
Since (\ref{35}) and (\ref{36}), the following inequality holds:
\vspace {3mm}
\beq                                          \label{37}
|G_1(x,\xi,t)| \leq  N(\varepsilon)   \varepsilon^{-\alpha}
e^{-qt\varepsilon}   +  N_1(\varepsilon) \ \varepsilon ^{-3/2} \
e^{-c^2t/\varepsilon^{2\alpha-1}}
\eeq
 \vspace{3mm}\noindent
where
\vspace {3mm}

\beq                                          \label{38}
N(\varepsilon) =
\frac{2\zeta(2)}{ql}[1-\varepsilon^{2(1-\alpha)}]^{-1/2} ; \ \ \
N_1(\varepsilon) =\frac{2 \zeta(2)
(2cl-\pi\varepsilon\beta)}{ql \sqrt{\pi \beta }
\sqrt{4cl-\beta \pi \varepsilon}}
\eeq

 \vspace{3mm}\noindent
and $\zeta(2)$ is the Riemann zeta function.

There remains to determine an upper bound for hyperbolic terms. This
may be done using inequalities proved in \cite{mda}.
So,beeing $\forall n \geq  [\bar k]+1$:
\vspace{3mm}
\beq                     \label{39}
bn^2t (1\pm \sqrt{1-(k/
n)^{2}}) \geq c^2 / \varepsilon,
\eeq
\vspace{3mm}\noindent
and since
\vspace{3mm}
\beq                          \label{310}
\sqrt{1-(k/n)^{2}} \geq \frac
{\pi \varepsilon (1-\beta)[4cl+\pi\varepsilon(1-\beta)]}{2cl+\pi\varepsilon
(1-\beta)},
\eeq
\vspace {3mm}\noindent
with $\beta \equiv 0$ if $k$ is an integer, we can write:
\vspace{3mm}
\beq           \label{311}
|G_2(x,\xi,t)|  \leq C_1(\varepsilon)  \ \varepsilon^{-2} \ e^{-c^2t/\varepsilon}
\eeq
\vspace{3mm}
\noindent
where
\vspace{3mm}
\beq                        \label{312}
C_1(\varepsilon)=\frac{2 \zeta(2)[cl+\pi\varepsilon
(1-\beta)]}{ql \pi  (1-\beta)[4cl+
\pi\varepsilon(1-\beta)]}.
\eeq
\vspace{3mm}

\noindent
The previous results lead to prove the following
\vspace{3mm}

\noindent
{\bf Theorem
3.1} {\em - The Green function} $G(x,\xi,t)  $ {\em defined in}
(\ref{26}) {\em converges absolutly for all }$(x,t)\in D$. {\em Moreover,
indicating by}$  \ M(\varepsilon)= max \{N_1(\varepsilon) \ \ \varepsilon
^{-3/2},C_1(\varepsilon) \ \ \varepsilon^{-2}\},$
{\em it holds}:
\vspace{3mm}
\beq                         \label{313}
|G(x,\xi,t)| \leq  N(\varepsilon)   \varepsilon^{-\alpha}
e^{-qt\varepsilon}   +  M(\varepsilon)
e^{-c^2t/\varepsilon^{2\alpha-1}}.
\eeq
\vspace{1cm}
 \section{Asymptotic approximation }
 \setcounter{equation}{0}

 \hspace{4.1mm}
Now, we are able to estimate function $r(x,t,\varepsilon)$ i.e. it is
possible to have an upper bound for the solution of problem
(\ref{23}).

\noindent In fact, recalling expression (\ref{24})-(\ref{25}), it
holds:
\vspace{3mm}
\beq                              \label{41}
| r(x,t,\varepsilon )| \leq l \varepsilon \int _0^t e^{-\varepsilon \tau}
\{ |\lambda_t(x,\tau)|+ \varepsilon |\lambda -u|\}
|G(x,\xi,t-\tau)|  d\tau  +
\eeq
\hspace* {2cm} \[+ l  \int_0^t |F(x,\tau)|
|1-e^{-\varepsilon \tau} |G(x,\xi,t-\tau)| d\tau.\]
\vspace{3mm}
\noindent
So, choosing:
\vspace {3mm}
\beq                        \label{42}
3/4<\alpha<1 \ \ \  and \ \ \  2(2\alpha-1)^{-1} <\delta<1,
\eeq
\vspace{3mm}\noindent
let:
\vspace{3mm}
\beq                      \label{43}
\beta =\delta (2\alpha-1)-1/2, \ \ \ \  0<\gamma<1.
\eeq
\vspace{3mm}\noindent
So, if
\vspace{3mm}\noindent
\beq             \label{44}
\ \ \ \ \eta= min\{\beta, \gamma, 1-\alpha, 1/2\} ;
\eeq
\vspace{3mm}\noindent
and
\vspace{3mm}
\beq
 \ \ A = max \{sup_{D} \  |F|,
sup_{D} \  |\lambda-u|, \ sup_{D} \ |\lambda_t|\}
\eeq
%\vspace {3mm}
%with
$$$$%\vspace{3mm}
%\beq                        \label{43}
%\beta =\delta (2\alpha-1)-1/2;\ \ \
%2(2\alpha-1)^{-1} <\delta<1
%\eeq
%\hspace*{2cm}\[3/4<\alpha<1; \ \ \  0<\gamma<1,\]
\vspace{3mm}\noindent
the following lemma holds:

{\bf Lemma 4.1 -} {\em If the function }$f(x,t,\varepsilon)$
{\em defined in} (\ref{24}) {\em is a continuous function in }$ D$ {\em
with continuous derivative with respect to }$ x$, {\em then the function} $r(x,t,\varepsilon)$ {\em
satisfies the inequality}
:
\vspace{3mm}
\beq                                         \label{44}
 |r| \leq  A l   \varepsilon ^\eta  \{ t^2
Z(\varepsilon)+
t Y(\varepsilon) +
\{t^{2-\delta}
+ t^{1-\delta}\} W(\varepsilon)
+ t^{1-\gamma} V(\varepsilon) \}+
\eeq
\hspace*{2cm}\[+ A \{U(\varepsilon)\ e^{-c^2t/\varepsilon} + S(\varepsilon)
\} \]

\vspace{3mm}\noindent
with
\vspace{3mm}
\beq                    \label{45}
Z(\varepsilon) = N(\varepsilon)/2; \ \ \  Y(\varepsilon)=max \{
2 N(\varepsilon), N_1(\varepsilon)\}
\eeq
\hspace*{2cm} \[ \  W(\varepsilon) = N_1(\varepsilon)
(\delta/e)^\delta \ \  \ \ V(\varepsilon) = C(\varepsilon)
[(1+\gamma)/e]^{1+\gamma} (1-\gamma)^{-1}\]
\vspace{1mm}
\[ \ U(\varepsilon)= 2 q \varepsilon  /c^2 \zeta(2)
+C(\varepsilon)/c^2+\varepsilon/c^2; \ \ \ S(\varepsilon)
= 2q \varepsilon \zeta(2)/c^2 + \varepsilon/c^2.\]

\vspace{3mm}
{\bf Proof}- Since the well known inequality \cite{m}:
\vspace{3mm}
\beq                           \label{46}
e^{-x} \leq [a/(ex)]^a    \ \ \   \forall a>0, \forall x>0
\eeq
\vspace{3mm}
\noindent
and
(\ref{41}),it holds:
\beq                         \label{47}
|r| \leq A l
[\varepsilon^{1-\alpha}(t^2/2+t+\varepsilon t) N +
N_1[\varepsilon^{\beta} (\delta/e)^\delta
(t^{2-\delta}+t^{1-\delta})+t \sqrt{\varepsilon}]+
\eeq
\vspace{3mm}
\hspace*{2cm}\[+C_1(\varepsilon) [{(1+\gamma)/e}^{1+\gamma}
(1-\gamma)^{-1} t ^{1-\gamma} \varepsilon^\gamma + \varepsilon/c^2
+e^{-c^2t/\varepsilon}(\varepsilon/c^2 + c^{-2})+\]

\vspace{1cm}\hspace*{1cm}
$+ 2\varepsilon \zeta(2) q /c^2  e^{-c^2t/\varepsilon}+  2q \varepsilon
/c^2 \zeta(2),$

\vspace {3mm}\noindent
from which, taking into account (\ref{42}), (\ref{43}), lemma follows.

In this way , if we consider the set
\vspace {3mm}
\beq                          \label {48}
Q_{\varepsilon} =\{ (x,t):  0\leq x \leq l , 0<t<\varepsilon^{-\eta/2}\}
\eeq
\vspace{3mm}\noindent
the following theorem holds:

{\bf Theorem 4.1 -} {\em When} $\varepsilon \rightarrow 0,${\em the solution
of the parabolic problem} (\ref{21}) {\em verifies the following
asymptotic estimate}
\vspace{3mm}
\beq                                         \label{49}
w(x,t,\varepsilon)= e^{-\varepsilon t}  u(x,t) + r(x,t,\varepsilon)
\eeq
\vspace{3mm}
\noindent
{\em where the error }$r(x,t,\varepsilon)$ {\em is uniformly bounded
every where in }$Q_{\varepsilon}.$

\end{document}